# Real-Time Head Gesture Recognition on Head-Mounted Displays using Cascaded Hidden Markov Models

Jingbo Zhao          Robert S. Allison

Department of Electrical Engineering and Computer Science
York University, Toronto, Canada
{jingbo, allison}@cse.yorku.ca

*Abstract*—Head gesture is a natural means of face-to-face communication between people but the recognition of head gestures in the context of virtual reality and use of head gesture as an interface for interacting with virtual avatars and virtual environments have been rarely investigated. In the current study, we present an approach for real-time head gesture recognition on head-mounted displays using Cascaded Hidden Markov Models. We conducted two experiments to evaluate our proposed approach. In experiment 1, we trained the Cascaded Hidden Markov Models and assessed the offline classification performance using collected head motion data. In experiment 2, we characterized the real-time performance of the approach by estimating the latency to recognize a head gesture with recorded real-time classification data. Our results show that the proposed approach is effective in recognizing head gestures. The method can be integrated into a virtual reality system as a head gesture interface for interacting with virtual worlds.

## I. Introduction

Many different interfaces have been developed for interactions between humans and virtual reality (VR) systems. interfaces [4], hand gesture interfaces [5], haptic interfaces [6] and olfactory displays [7]. Although head gesture is a natural and important way for people to communicate and interact with each other, recognition of head gestures in the context of virtual reality systems and the use of head gesture recognition to interact with avatars and virtual environments have been largely ignored in previous VR related research. In the current study, we present an approach for recognizing head gestures using a head-mounted display (HMD).

Enabling real-time head gesture recognition on HMDs could be useful. For instance, in virtual reality systems, users usually need to interact with avatars. To answer Yes/No questions asked by avatars, users could simply make their responses by nodding and shaking heads through a head gesture interface. One possible application is to use the interface to interact with virtual tour guides in augment reality (AR) based tours [8]. It is also very common that a virtual reality system itself may raise questions to users and ask users to confirm or reject certain options; in this case, the user can also respond through a head gesture interface by nodding and shaking. Recently, there is a growing interest for teleoperation of robots, such as quadcopters, using head motions tracked by HMDs [9-12]. First-person views of robots are often presented directly through the display panels of HMDs. In the case of the quadcopter control, a quadcopter can be maneuvered by head spatial translations [11] or the head orientation can be used to manipulate the attitude of the quadcopter to fly it [12]. Adding a head gesture interface in such applications will enable users to perform more complex operations. For example, users could nod their head to perform mode switching to switch flight control from auto-pilot to head motion control. Head gesture interfaces also can be applied to (AR) devices, such as the Google Glass and the Microsoft HoloLens. Such an interface would enable users to perform actions, such as browsing, with head rotation, head tilting, nodding and shaking without touching the glasses. A head gesture recognition method is beneficial to researchers who investigate human locomotion or driving behaviors etc. In such activities, users usually have to make head movements for observing the environment around them. A head gesture recognition method can be used to count the number, the type and the duration of each head movement. These parameters may differ significantly given different experimental conditions, such as restricted fields of view and complexity of the environment. Without such an approach for recognizing head gestures, researchers have to manually determine the number, the type and the duration of each head movement from the collected data of head movements. Lastly, a head recognition module also can be integrated to driver assistance systems of automobiles to monitor the driving behaviors of drivers [13].

Recognizing gestures is usually considered as the problem of recognizing sequences and HMMs have been widely used for recognizing hand and body gestures [14-17]. Previous studies on head gesture recognitions were mostly computer vision based systems using HMMs [18],[19]. In such systems, a user's face was usually captured by a camera and computer vision algorithms were used to track a user's face and estimate the user's head orientation from the tracked face.

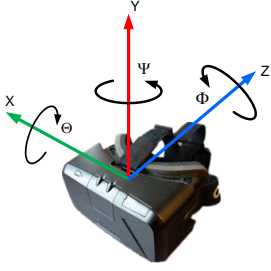

Fig. 1. The coordinate system of the Oculus Rift DK2 [20].

Morimoto et al. [18] used an optical flow algorithm to estimate the yaw, pitch and roll of a user's head. The estimated angles were quantified into seven observation symbols by thresholding. The thresholds were determined by calculating the energy of a sequence of angles, but the details of the energy calculation algorithm were not described. The observation symbols were used to train four HMMs. These HMMs correspond to four gestures: Yes, No, Maybe and Hello. Three gestures Yes, No and Maybe were modeled with fully connected HMMs while the gesture Hello was modeled with a left-right HMM. The classification performance was evaluated offline using sequences of gestures collected from several participants. Terven et al. [19] used the Supervised Descent Method (SDM) to extract 2-D facial features. The yaw, pitch and roll were estimated by using the POSIT (Pose from Orthography and Scaling with Iterations) method with the extracted 2-D facial features and a 3-D anthropometric head model. The symbols were generated by comparing changes in yaw and pitch across consecutive frames. To cover six different head gestures: Nodding, Shaking, Left, Right, Up and Down, the researchers trained six HMMs and these models were fully connected to form a cyclic structure. To evaluate the classification performance, videos that contained different head gestures were recorded and the performance was evaluated offline.

Although computer vision based systems using HMMs for head gesture recognition exist, to the authors' knowledge, HMMs have not been used for recognizing head gestures on HMDs. In this paper, we present the Cascaded Hidden Markov Models (CHMMs) for head gesture recognition on HMDs. Unlike existing systems that contain cyclic structure, our system is fully pipelined, making it more suitable for real-time applications.

The main contributions of the paper are:

1) We proposed the CHMM structure for real-time head gesture recognition. The system recognizes nine types of head gestures and can be further extended to include more head gestures;

2) We proposed a training method that maximizes the classification performance of the CHMM;

3) We presented a procedure for simultaneously capturing and labeling head gestures using HMDs;

The present study focused on a specific HMD -- the Oculus Rift DK2, but the proposed method is general and can be adapted to work with VR systems that use head tracking glasses or other types of systems in which a user's head motion is tracked by fast and accurate tracking systems.

## II. METHODS

### A. The Tracking System of the Oculus Rift DK2

The Oculus Rift DK2 uses a six Degree-of-Freedom (DOF) hybrid optical-inertial tracker to track a user's head motion at approximately 75 Hz. The coordinate system of the Oculus Rift DK2 is illustrated in Fig. 1. The hybrid tracker consists of an external camera with an infrared filter to track the infrared LED array on the front and side panels of the Oculus Rift DK2 and an embedded inertial measurement unit (IMU) [21],[22]. The tracking data that can be accessed through the VR software WorldViz Vizard 5.0 are position, acceleration, Euler angles and angular velocities. To recognize head gestures, we only used the angular velocities of head motions as angular velocities directly reflect whether a user's head is moving and in which direction. We represent the angular velocity as a 3-D vector $\omega = (\dot{\Psi}, \dot{\Theta}, \dot{\Phi})$, where $\dot{\Psi}, \dot{\Theta}, \dot{\Phi}$ are yaw velocity, pitch velocity and roll velocity, respectively. A sequence of angular velocities can be denoted as $W = \omega_1 \omega_2 \ldots \omega_i$, where $i$ is the index for a 3-D vector $\omega$. Another option for monitoring head movements is to use quaternions to represent the head angular velocity but extra time will be taken to convert head angular velocities to quaternions.

### B. Definition of Head Gestures

We defined nine classes of head gestures. Seven are simple gestures: Being Idle (remaining still), Rotating Left, Rotating Right, Tilting Upward, Tilting Downward, Leaning Left, Leaning Right. Two are complex head gestures: Shaking, Nodding.

The motivation behind the definition of complex gestures is that shaking can be represented a sequence of three simple head gestures, which are Being Idle, Rotating Left and Rotating Right. Similarly, Nodding can be represented by a sequence of simple gestures: Being Idle, Tilting Upward and Tilting Downward. We associated each class of head gesture with a class label $l$, with $l \in \{1,2,\ldots,9\}$.

### C. Cascaded Hidden Markov Models

An HMM [23],[24] is governed by the following parameters: $N$ the number of hidden states, $M$ the number of observation symbols and the model parameter $\lambda = (A, B, \pi)$, where $A$ is the matrix that represents the transition probability between states, $B$ the matrix that represents the emission probability of a symbol observed from a specific state and $\pi$ the initial state probabilities. Similar to the speech recognition approach [23], we modeled each head gesture with a left-right HMM $\lambda_l$, where $l$ is the class label of a head gesture associated with the HMM $\lambda$. In a left-right HMM, only transitions between adjacent states from left to right and transitions from a state to itself are allowed. A set of trained HMMs for the nine classes of head gestures can be represented as $\Lambda$, $\Lambda = \{\lambda_1, \lambda_2, \ldots, \lambda_9\}$. The HMMs we used were discrete HMMs and discrete HMMs only accept discrete observation symbols as inputs. Thus, given an sequence of sampled angular velocities $W$, before feeding the sequence $W$ into an HMM $\lambda_l$, we used the K-Means algorithm during training step and the minimum distance classifier [25] during testing step as the vector quantization (VQ) procedure to quantize the sequence of angular velocities $W$ into an observation sequence $S$ that consists of discrete observation symbols with $S = O_1 O_2 \ldots O_i$, where $i$ is the index of the observation symbol $O$ in the sequence $S$. To predict how likely a sequence $S$ belongs to a certain class $l$ of a head gesture, we used a set of trained HMMs $\Lambda$ and the forward procedure of the HMM to calculate the posterior probabilities $P_a$ for all HMMs, with $P_a = \{ P(S \mid \lambda_1), P(S \mid \lambda_2), \ldots, P(S \mid \lambda_9) \}$. An output selection procedure resolves the class label $l$ of the given observation sequence $S$ from the posterior probabilities $P_a$. As simple gestures require much less data to give a reliable estimate than complex gestures do, to efficiently recognize simple and complex head gestures, we organized the set of HMMs $\Lambda$ into the CHMM structure. The structure has two dedicated layers for recognizing simple gestures and complex gestures respectively. For training and testing the CHMM, two HMM algorithms were used. One is the Baum–Welch algorithm, which was used to train a left-right HMM $\lambda_l$. The other is the forward procedure of the HMM, which calculates the posterior probability $P(S \mid \lambda_l)$ of a HMM $\lambda_l$ given an observation sequence $S$. These two algorithms are available in Matlab 2014a as hmmtrain() and hmmdecode(). Detailed descriptions of the algorithms can be found in [23],[24].

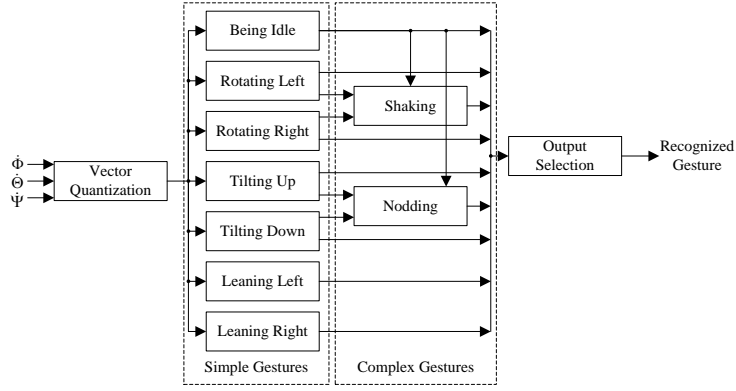

Fig. 2. The structure of the CHMMs for real-time head gesture recognition.

In Fig. 2, we present the proposed CHMM structure. Here we describe the real-time operation of the CHMM and leave the explanation of the training and testing procedures to Section 3. During real-time operation, the system continuously reads angular velocity $\omega_i$ for processing. A sampled angular velocity $\omega_i$ is given into the vector quantization module and the vector quantization module produces an observation symbol $O_i$ based on the Euclidean distance between cluster centers $C_j$ and the vector $\omega_i$ using the minimum distance classifier [25]: the index $j$ of cluster center $C_j$ that gives the shortest Euclidean distance is assigned as the observation symbol to the vector $\omega_i$:

$$O_i = \underset{j}{\operatorname{argmin}} \left( \left\| C_j - \omega_i \right\|_2 \right) \quad (1)$$

where $O_i$ is the assigned observation symbol and cluster centers $C_j$ were obtained by K-Means during training. The observation symbol $O_i$ generated by the vector quantization module is buffered using an array until the number of observation symbols $O_i$ reaches the length $L_B$. When the number of observation symbol $O_i$ reaches $L_B$, the sequence of observation symbols $S$ is sent to the modules classifying simple gestures to calculate the posterior probabilities $P_s$, with $P_s = \{ P(S \mid \lambda_1), P(S \mid \lambda_2), \ldots, P(S \mid \lambda_7) \}$. The buffer is then immediately cleared and waits for new observation symbols $O_i$. If a recognized simple gesture $l$ is in the category of Being Idle, Rotating Left, Rotating Right, Tilting Up or Tilting Down, they will be considered as observation symbols $O_i'$ for the layer of complex gestures and will be further buffered using a queue of a length $L_Q$. Each time the queue performs a dequeue and an enqueue operation, the buffered sequence in the queue will be sent to the module of complex gestures to calculate the posterior probabilities $P_c$ of complex gestures, with $P_c = \{ P(S \mid \lambda_8), P(S \mid \lambda_9) \}$. We empirically set $L_B = 10$ and $L_Q = 10$ to make the array contain 0.13 s of data and the queue contain 1.3 s of data at the sampling rate of 75 Hz. We found such choice of values gave a relatively fast response for recognizing simple gestures and reliable length of data for recognizing complex gestures. The last step is to resolve the class label $l$ of the head gesture based on the calculated posterior probabilities $P_s$ and $P_c$. The posterior probabilities of simple gestures $P_s$ and the posterior probabilities of complex gestures $P_c$ are not directly comparable as a complex gesture consists of symbols represented by simple gestures. To solve this problem, we proposed an output selection procedure to estimate the class label $\hat{l}$ from the posterior probabilities of simple gestures $P_s$ and complex gestures $P_c$ using the function:

$$\hat{l} = \begin{cases} \underset{l}{\operatorname{argmax}}(P_c) & \text{if } P(S \mid \lambda_8) > \tau_n \text{ or} \\ & \text{if } P(S \mid \lambda_9) > \tau_s \\ \underset{l}{\operatorname{argmax}}(P_s) & \text{otherwise} \end{cases} \quad (2)$$

where $\tau_n$ and $\tau_s$ are the thresholds that indicate a complex gesture shaking or nodding may exist if either the posterior probability $P(S \mid \lambda_8)$ or $P(S \mid \lambda_9)$ is larger than their corresponding thresholds $\tau_n$ or $\tau_s$; $l$ is the class label associated with a HMM $\lambda$ of a specific gesture and $\hat{l}$ is the estimated class label.

TABLE I. THE AVERAGE ACCURACY OF THE SIMPLE GESTURE LAYER FROM A TRAINING SESSION (UNIT: PERCENTAGE, M: THE NUMBER OF DISCRETE SYMBOLS IN HMMS, N: THE NUMBER OF HIDDEN STATES IN HMMS).

| | | \multicolumn{19}{c}{M} |
|---|---|---|---|---|---|---|---|---|---|---|---|---|---|---|---|---|---|---|---|
| | | 7 | 8 | 9 | 10 | 11 | 12 | 13 | 14 | 15 | 16 | 17 | 18 | 19 | 20 | 21 | 22 | 23 | 24 | 25 |
| | 2 | 87.6 | 90.1 | 95.9 | 92.6 | 95.2 | 97.2 | 97.5 | 98.6 | 97.0 | 98.7 | 98.6 | 98.9 | 97.6 | 98.7 | 98.8 | 97.6 | 97.4 | 97.5 | 97.4 |
| | 3 | 85.7 | 90.1 | 95.9 | 91.6 | 97.0 | 96.2 | 98.8 | 98.7 | 99.0 | 98.9 | **99.2** | 98.4 | 98.9 | 99.1 | 97.4 | 99.0 | 97.5 | 97.6 | 97.5 |
| N | 4 | 87.6 | 92.7 | 95.9 | 97.4 | 97.0 | 97.3 | 99.0 | 98.4 | 99.1 | 98.8 | 99.0 | 99.1 | 99.0 | 97.3 | 97.5 | 97.2 | 97.3 | 97.7 | 97.8 |
| | 5 | 86.1 | 90.3 | 95.9 | 94.0 | 97.4 | 98.6 | 98.7 | 98.7 | 98.8 | 99.0 | 99.0 | 99.0 | 98.9 | 97.5 | 99.0 | 98.8 | 97.8 | 97.4 | 97.6 |
| | 6 | 88.1 | 90.4 | 96.2 | 96.1 | 98.4 | 98.7 | 98.7 | 98.3 | 98.7 | 99.1 | 99.1 | 98.9 | 98.9 | 99.0 | 98.9 | 98.7 | 97.4 | 97.0 | 97.2 |

## III. EXPERIMENT 1: TRAINING AND TESTING OF THE CHMM

In experiment 1, we trained the CHMM and evaluated its offline classification performance. As there was no publicly available head gesture dataset for the Oculus Rift DK2, we developed a custom application, using the Vizard 5.0, that can simultaneously collect and label head gestures for training and testing the proposed CHMM structure. Nineteen people participated in the experiment and informed consents were obtained from all participants in accordance with a protocol approved by the Human Participants Review Subcommittee at York University. In the experiment, the participants wore the Oculus Rift DK2 and sat approximately 60 cm in front of the tracking camera of the Oculus Rift DK2, which was mounted on the monitor of the host machine (Windows 7, an Intel i7 2.8 GHz. CPU, 4 GB memory and an AMD Radeon HD 6850 graphics card). There were nine types (classes) of head gestures that needed to be collected. For each type of head gesture, the researcher pressed the corresponding button on the control panel (visible only to the researcher on the computer monitor) of the custom application and a prompt (visible to both participants on the HMD and the researcher on the computer monitor) indicating the type of head gesture that a participant needed to perform was shown in the view of the HMD. A countdown timer was also started at the same time to count from two to zero by seconds. A participant was expected to complete the head gesture within 2 seconds with their preferred head movement speed. Labeling of the head gesture was done at the same time by the custom application within the 2-second interval. We collected each type of head gesture twice for each participant. In total, we had 342 head gesture samples from nineteen participants.

Our training and testing procedures were fully automated and performed using Matlab 2014a. We used multi-class precision ($Precision_M$), multi-class recall ($Recall_M$) and the average accuracy [26] as the metrics to evaluate the performance of training and testing. We divided the training of the CHMM into three phases that trained the modules of vector quantization, simple gesture layer and complex gesture layer separately. Given the collected head gesture dataset from nineteen participants, we directly ran the K-means algorithm to quantize each angular velocity vector $\omega_i$ into an observation symbol $O_i$. The cluster centers $C_j$ obtained from K-Means training were stored for the testing step for the vector quantization module. The cluster number $K$ of K-Means and the number of observation symbols $M$ of HMMs were equated ($M = K$). Seven types of observation symbols, with $M \geq 7$, were needed to represent seven simple gestures. The head gesture dataset was then divided in halves into a training set and a testing set. For each sequence of head gestures other than being idle, we removed redundant symbols that indicate a user's head is remaining still as the redundant symbols were not useful for training.

The second step was to train the layer of simple gestures and evaluate its classification performance. To represent a simple head gesture, such as rotating left, at least two hidden states $N$ are needed, with $N \geq 2$. The Baum–Welch algorithm was used for training. Since we chose $L_B = 10$ for real-time evaluation, we partitioned each sequence in the training set into short sequences of length 10 and used the short sequences to train its each associated HMM $\lambda_l$, $l \in \{1, 2, ..., 7\}$, to obtain the parameters of the transition matrix $A$ and the emission matrix $B$. The initial state probabilities $\pi$ were not considered as a head gesture always starts with the state of the head being remaining still. The information can be learned during training and stored in the emission matrix $B$. There were two tunable parameters: $N$ the number of hidden states in the model, $M$ the number of observation symbols in an HMM and we knew that $N \geq 2$ and $M \geq 7$. To obtain the best classification performance, we wished to find the optimal values of $N$ and $M$ that maximize the average accuracy for the layer of simple gestures, with smallest $M$ and $N$ possible. The smaller $N$ and $M$ are, fewer additions and multiplications are involved in the forward procedure of the HMM; hence the faster real-time performance for head gesture recognition. The upper bounds for $N$ and $M$ were set to 6 and 25 empirically. The evaluation of the classification performance of the simple gesture layer was conducted after each training cycle with a combination of $N$ and $M$. We partitioned each sequence in the testing set into the short sequences of length 10 and used the forward procedure to calculate the posterior probabilities $P_s$ based on short sequences. The class label $\hat{l}$ of a simple head gesture was estimated using the equation:

$$\hat{l} = \underset{l}{\mathrm{argmax}}(P_s) \qquad (3)$$

Since the K-means is initialized randomly, the training results may differ even if we run the same algorithm on the same training set. Thus, we ran the training procedure for the layer of simple gestures for five different sessions. In each session, the training was performed with different combinations of $N$ and $M$ such that $2 \leq N \leq 6, N \in \mathbb{Z}$ and $7 \leq M \leq 25, M \in \mathbb{Z}$. The highest average accuracy we were able to obtain from a specific training session was 99.2% when $N = 3$ and $M = 17$ (see Table I) and the corresponding multi-class precision and multi-class recall were 97.9% and 96.6%, respectively.

TABLE II. HEAD GESTURE RECOGNITION LATENCIES (UNIT: S).

|      | RL    | RR    | TU    | TD    | LL    | LR    | S     | N     |
|------|-------|-------|-------|-------|-------|-------|-------|-------|
| P1   | 0.213 | 0.253 | 0.173 | 0.120 | 0.200 | 0.240 | 0.720 | 0.653 |
| P2   | 0.173 | 0.173 | 0.147 | 0.120 | 0.107 | 0.160 | 0.747 | 0.733 |
| P3   | 0.107 | 0.093 | 0.133 | 0.173 | 0.227 | 0.253 | 0.813 | 0.667 |
| P4   | 0.267 | 0.080 | 0.093 | 0.200 | 0.253 | 0.120 | 0.773 | 0.680 |
| P5   | 0.280 | 0.280 | 0.173 | 0.147 | 0.080 | 0.173 | 0.667 | 0.733 |
| P6   | 0.267 | 0.133 | 0.160 | 0.200 | 0.120 | 0.120 | 0.693 | 0.693 |
| Mean | 0.218 | 0.169 | 0.147 | 0.160 | 0.164 | 0.178 | 0.736 | 0.693 |
| Std  | 0.068 | 0.083 | 0.030 | 0.037 | 0.071 | 0.058 | 0.054 | 0.034 |

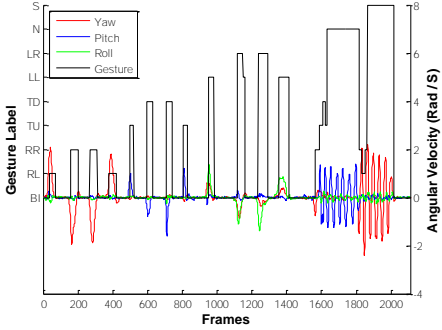

Fig. 3. An exemplar real-time recognition result from a participant.

To train the complex gesture layer and evaluate the classification performance, we first partitioned the sequence of nodding and shaking in the head gesture datasets into short sequences of length 10 (since $L_Q = 10$). We then used the trained HMMs $\Lambda_s = \{\lambda_1, \lambda_2, ..., \lambda_7\}$ of simple gestures to classify the short sequences of complex gestures into observation symbols $O_i'$ that consists of classified simple gestures. Specifically, the classification was done by first using the forward procedure of the HMM to calculate the posterior probabilities of each short sequence of complex gestures and assigning the short sequence with the class label of the simple gesture with the highest posterior probability using equation (3). As a complex gesture is represented by three simple gestures, we set the number of observation symbols $M = 3$. The number of hidden states was set as $N = 3$, which was same as that of the simple gesture layer. We then used the Baum–Welch algorithm to train the layer of complex gestures and obtained the parameters for the transition matrix $A$ and the emission matrix $B$ for HMMs of complex gestures $\Lambda_c = \{\lambda_8, \lambda_9\}$. As with the training of the simple gesture layer, the initial state probabilities $\pi$ were not considered. To determine the thresholds $\tau_n$ and $\tau_s$, we used the forward procedure and the HMMs $\Lambda_c = \{\lambda_8, \lambda_9\}$ to calculate the posterior probabilities $P_c$ of all sequences of complex head gestures in the training set. We selected the smallest values as the thresholds $\tau_n$ and $\tau_s$ for shaking and nodding, respectively, with $\tau_n$ = -6.93 and $\tau_s$ = -7.54. The last step was to test the layer of complex gestures. We calculated the posterior probability $P_c$, $P_c = \{P(S|\lambda_8), P(S|\lambda_9)\}$, of each complex gesture sequence in the testing set, compared the posterior probabilities $P_c$ with the thresholds $\tau_n$ and $\tau_s$ we obtained during the training procedure and estimated the class label of each sequence using the equation:

$$\hat{l} = \begin{cases} \underset{l}{\mathrm{argmax}}(P_c) & if \ P(S|\lambda_8) > \tau_n \ or \\ & if \ P(S|\lambda_9) > \tau_s \\ -1 & otherwise \end{cases} \quad (4)$$

where -1 will be given as an invalid class label when both $P(S|\lambda_8)$ and $P(S|\lambda_9)$ were lower than their corresponding thresholds $\tau_n$ and $\tau_s$.

The multi-class precision, the multi-class recall and the average accuracy were 100%, 96.4% and 98.5%, respectively, for the layer of complex gestures.

IV. EXPERIMENT 2: REAL-TIME EVALUATION OF THE LATENCY OF THE CHMM

For real-time evaluation of the latency of the head gesture recognition framework, we implemented the proposed CHMM structure and the forward procedure of the HMM using python 2.7 in Vizard 5.0. Our goal was to estimate the latency for the algorithm to recognize a head gesture. In practice, we found it necessary to further tune the parameters $\tau_n$ and $\tau_s$ such that $\tau_n = -5$ and $\tau_s = -4$. This helped the CHMM avoid confusing fast head rotation and tilting with Shaking and Nodding during real-time recognition. Each computation cycle of the CHMM takes approximately 1 ms and thus the proposed approach meets the real-time requirement of completing a computation cycle within 13 ms at the sampling rate of 75 Hz. Nine people participated in the experiment and none had participated in experiment 1. Informed consents were obtained from all participants in accordance with a protocol approved by the Human Participants Review Subcommittee at York University. In experiment 2, participants also wore the Oculus Rift DK2 and sat 60 cm in front of the tracking camera of the Oculus Rift DK2. Participants were asked to perform the following head gesture sequence with their preferred speed three times from the initial neutral position with a head gesture of Being Idle (BI):

1) Rotating Left (RL);
2) Move Back to Neutral Position;
3) Rotating Right (RR);
4) Move Back to Neutral Position;
5) Tilting Upward (TU);
6) Move Back to Neutral Position;
7) Tilting Downward (TD);
8) Move Back to Neutral Position;
9) Leaning Left (LL);
10) Move Back to Neutral Position;

11) Leaning Right (LR);
12) Move Back to Neutral Position;
13) Nodding (N);
14) Move Back to Neutral Position;
15) Shaking (S);
16) Move Back to Neutral Position;

The estimated head gesture labels and head angular velocities during real-time operation were recorded for latency analysis (see Fig. 3). We defined the latency for head gestures, except Being Idle, as the time interval between the index of the frame $i_{init}$ when the Euclidean norm of a user's head angular velocity $\|\omega\|$ is equal to or higher than a threshold $\omega_{init} = 0.1$ rad / s, to the index of the frame $i_{trigger}$ at which a head gesture was triggered. Then the latency $t_L$ can be estimated as:

$$t_L = \frac{i_{trigger} - i_{init}}{f_s} \quad (5)$$

where $f_s$ is the sampling frequency of the Oculus Rift DK2 and $f_s = 75$ Hz.

Three participants were unable to follow the gesture sequence given above as they did not remember the sequence they needed to perform. Thus, we only used the data of the remaining six participants and the researcher selected one specific sequence among the three that a participant performed to estimate the latency. The selected sequence was the clearest pattern compared with that of other two sequences. In table II, we present the result of the estimated head gesture recognition latencies of six participants (P1 – P6). The mean value and the standard deviation of the latencies of all participants were also calculated. We can find that for simple gestures, the latency had a mean value of 0.17 s and for complex gestures the latency had a mean value of 0.71 s.

## V. CONCLUSIONS

We have presented a CHMM structure for real-time head gesture recognition. The proposed structure is scalable and modules for other types of gestures can be added or the existing modules can be removed based on the application needs. A distinct advantage of the proposed pipelined structure is that the structure can be more easily implemented on FPGAs (Field Programmable Gate Arrays) and ASICs (Application-specific Integrated Circuits) compared with the cyclic structure described in [19], as a cyclic structure is iterative and non-deterministic. Such modules can be integrated into head wearable devices, such as the Google Glass and Microsoft HoloLens, as a dedicated module for fast head gesture recognition. A limitation with the user-study based real-time evaluation is that it is impossible to ask the user to perform head motions with precise velocity and duration. Thus, the real-time classification performance of the proposed approach needs to be further evaluated with a robotic head since the velocity and the duration of a head gesture performed by a robotic head can be easily controlled by programming. This will enable us to compare the timings of head movements with that of the real-time classification results and determine the classification performance.

## REFERENCES

[1] C. Ware, K. Arthur and K. S. Booth. Fish tank virtual reality. In *Proc. INTERACT 1993 and CHI 1993*. 37-42.

[2] I. Sutherland. A head-mounted three dimensional display. In *Proc. FJCC 1968*. 757-764.

[3] C. Cruz-Neira, D. J. Sandin and T. A. DeFanti. Surround-Screen Projection-Based Virtual Reality: The Design and Implementation of the CAVE. In *Proc. SIGGRAPH 1993*.

[4] J.M. Hollerbach. Locomotion interfaces. In *Handbk. of VE: Dsg. Impl. and App.*, Lawrence Erlbaum Associates, Inc., 239-254. 2002.

[5] D. Xu. A Neural Network Approach for Hand Gesture Recognition in Virtual Reality Driving Training System of SPG. In *Proc. ICPR 2006*. 519-522.

[6] I. Choi and S. Follmer. Wolverine: A Wearable Haptic Interface for Grasping in VR. In *Proc. UIST 2016*. 117-119.

[7] H. Matsukura, H. Yoshida, T. Nakamoto and H. Ishida. Synchronized presentation of odor with airflow using olfactory display. In *Jrnl. Mech. Sci. Tech.* 24, 1, 253-256. 2010.

[8] A.F. Abate, G. Acampora and S. Ricciardi. An interactive virtual guide for the AR based visit of archaeological sites. In *Jrnl. Vi. Lang. & Compu.* 22, 6, 415-425. 2011.

[9] H. Martins and R. Ventura. Immersive 3-d teleoperation of a search and rescue robot using a head-mounted display. In *IEEE Conf. ETFA 2009.* 1–8.

[10] N. Mollet and R. Chellali. Virtual and augmented reality with head-tracking for efficient teleoperation of groups of robots. In *IEEE Conf. Cyberworlds 2008*, 102–108.

[11] K. Higuchi, K. Fujii and J. Rekimoto. Flying head: A head-synchronization mechanism for flying telepresence. In *Proc. ICAT 2013*, 28-34.

[12] C. Pittman and J. J. LaViola, Jr.. Exploring head tracked head mounted displays for first person robot teleoperation. In *Proc. IUI 2014*. 323-328.

[13] H. B. Kang. Various Approaches for Driver and Driving Behavior Monitoring: A Review. In *IEEE ICCVW 2013*. 616-623.

[14] K. Liu, C. Chen, R. Jafari and N. Kehtarnavaz. Multi-HMM classification for hand gesture recognition using two differing modality sensors. In *Proc. IEEE DCAS 2014*, 1-4.

[15] M. Hossain and M. Jenkin. Recognizing hand-raising gestures using HMM. In *Proc. CRV 2005*, 405-412.

[16] L. W. Campbell, D. A. Becker. A. Azarbayejani, A. F. Bobick, and A. Pentland. Invariant features for 3-D gesture recognition. In *Proc. FG 1996*, 157.

[17] C. Chen, J. Liang, H. Zhao, H. Hu and J. Tian. Factorial HMM and Parallel HMM for Gait Recognition. In *IEEE Trans. Syst. Man. Cybern. C (App. and Rev.)*, 39, 1, 114-123. 2009.

[18] C. Morimoto, Y. Yacoob and L. Davis. Recognition of head gestures using hidden Markov models. In *Proc. ICPR 1996*. 3, 461-465.

[19] J. R. Terven, J. Salas and B. Raducanu. Robust Head Gestures Recognition for Assistive Technology. In *Proc. MCPR 2014*, 152-161.

[20] https://developer3.oculus.com/documentation/pcsdk/latest/concepts/dg-sensor/, last accessed Dec 2016.

[21] S. M. LaValle, A. Yershova, M. Katsev and M. Antonov. Head tracking for the Oculus Rift. In *Proc. ICRA 2014*, pp. 187-194.

[22] Hacking the Oculus Rift DK2. http://doc-ok.org/?p=1095, last accessed Dec 2016.

[23] L. R. Rabiner. A tutorial on hidden Markov models and selected applications in speech recognition. In *Proc. IEEE*, 77, 2, 257-286. 1989.

[24] D. Ramage. Hidden Markov models fundamentals. Lecture Notes. 2007. http://cs229.stanford.edu/section/cs229-hmm.pdf, last accessed Dec 2016.

[25] H. Lin and A. N. Venetsanopoulos. A weighted minimum distance classifier for pattern recognition, In *Proc. Canadian Conf. on ECE*, 2, 904-907. 1993.

[26] M. Sokolova and G. Lapalme. A systematic analysis of performance measures for classification tasks. *Inf. Process. Manage*. 45, 4, 427-437. 2009.